\documentclass{SCTYE}
\usepackage{multicol}
\begin{document}

\begin{picture}(0,0){\rm
\put(0,4){\makebox[160truemm][l]{\sf  
Basic Plasma Processes in Solar-Terrestrial Activities}}}
\end{picture}

\begin{picture}(0,0){\rm
\put(0,36){\makebox[178truemm][l]{\textcolor[rgb]{0.39,0.39,0.39}{\xiaosihao
\sf Article\hfill
\includegraphics{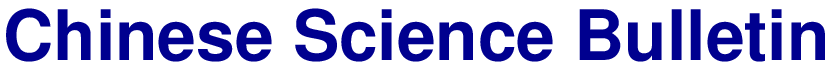}}}}}
\end{picture}

\def\bm{\boldsymbol}

\def\dl{\displaystyle}
\def\wyl{\end{multicols}\end{document}}
\newcommand{\pratio}{\sigma}

\Year{2011} %
\Month{October}
\Vol{56} %
\No{1} %
\BeginPage{1} %
\EndPage{4} %
\AuthorMark{{{\rm Chen, P. F.,} et al.}}
\DOI{10.1007/s11434-011-4829-9} 

\title{Where do flare ribbons stop?}

\author[1,2]{Chen, P. F.}{Corresponding author (email: chenpf@nju.edu.cn)}
\author[2]{Su, J. T.}{}
\author[1]{Guo, Y.}{}%
\author[2]{Deng, YuanYong}{}

\address[{\rm1}]{Department of Astronomy, Nanjing University, Nanjing 210093, China;}
\address[{\rm2}]{Key Laboratory of Solar Physics, National Astronomical Observatories, Chinese Academy of Sciences, Beijing 100012, China}

\maketitle \vspace{-2mm}{\footnotesize Received August 31, 2011; accepted September 19, 2011
}\vspace*{3mm}


\rule{16.8cm}{0.4pt}\vspace{1mm}\\
\parbox{16.8cm}
{\begin{abstract}
The standard flare model, which was proposed based on observations and
magnetohydrodynamic theory, can successfully explain many observational features
of solar flares. However, this model is just a framework, with many details
awaiting to be filled in, including how reconnection is triggered. In this
paper, we address an unanswered question: where do flare ribbons stop? With the
data analysis of the 2003 May 29 flare event, we tentatively confirmed our
conjecture that flare ribbons finally stop at the intersection of separatrices
(or quasi-separatrix layer in a general case) with the solar surface. Once
verified, such a conjecture can be used to predict the final size and even the
lifetime of solar flares.
\end{abstract}}

\vspace*{0.35cm} \noindent
\parbox{16.8cm}{\keywords{solar flares, magnetic field, separatrix}}
\vspace{3mm}

\rule{16.8cm}{0.4pt}\vspace{-0.8mm}\\
\renewcommand{\baselinestretch}{1.2}
\renewcommand{\arraystretch}{1.5}
{\begin{tabular}{lp{0.88\textwidth}}  \scriptsize
{\bf\hspace{-2.6mm} Before publication, please cite the paper as:}\!\!\!\!&\scriptsize Chen, P. F., Su, J. T., Guo,
Y., \& Deng, Y. Y. 2011, Chinese Science Bulletin, 2011, in
press, doi:10.1007/s11434-011-4829-9
\\
\end{tabular}}

\rule{16.8cm}{0.4pt}


\textwidth=178truemm \textheight=236truemm

\wuhao\vspace*{5mm}
\begin{multicols}{2}

\renewcommand{\baselinestretch}{1.08} \baselineskip 12.2pt\parindent=10.8pt

\renewcommand{\thefootnote}

\noindent Owing to the omnipresence of magnetic field in the solar atmosphere,
the Sun presents a variety of activities, which are modulated with the 11-year
solar cycles. One of the spectacular phenomena is solar flares. They represent
the typical process that magnetic energy accumulated gradually in the corona is
converted rapidly into thermal and kinetic energies.
Morphologically, flares are classified into two types, i.e., compact and
two-ribbon flares (e.g., \cite{pall91}). Compact flares are characterized by a
compact flare loop, which does not show significant change in shape, whereas
two-ribbon flares are characterized by flaring loop expansion and bright ribbon
separation. Two-ribbon flares attracted more attention since they are frequently
related to coronal mass ejections (CMEs).

To explain the appearance of the flaring loops, two ribbons, and their 
association with filament eruptions, a standard flare model was gradually
developed by Carmichael \cite{carm64}, Sturrock \cite{stur66}, Hirayama 
\cite{hira74}, and Kopp \& Pneuman \cite{kopp76}, which was later called CSHKP
model. The standard flare model, where magnetic
reconnection below an erupting flux rope is the key ingredient, was supported
by a lot of observations, such as the discoveries of chromospheric evaporation,
the cusp-shaped structure, the reconnection downflow, and inflow (see 
\cite{huds11} for a review). However, it should be kept in mind that such a
model is just a framework, and many detailed processes inside it await to be
clarified and understood theoretically, for example, how the reconnection is
triggered in the highly-conducting plasma. Another unclarified issue is
related to the flare ribbon separation.

The typical feature of two-ribbon flares in H$\alpha$ or any other
chromospheric wavelength is the ribbon separation. The separating speed reaches
up to 50 km s$^{-1}$ in the impulsive phase and decreases to $\leq 1$ km 
s$^{-1}$ in the decay phase \cite{nolt79}. One important question remaining 
unanswered is: where do the two ribbons finally stop? This paper is aimed to
address such an issue.

\section{Our conjecture}

\end{multicols}

\centerline{\includegraphics[width=14cm]{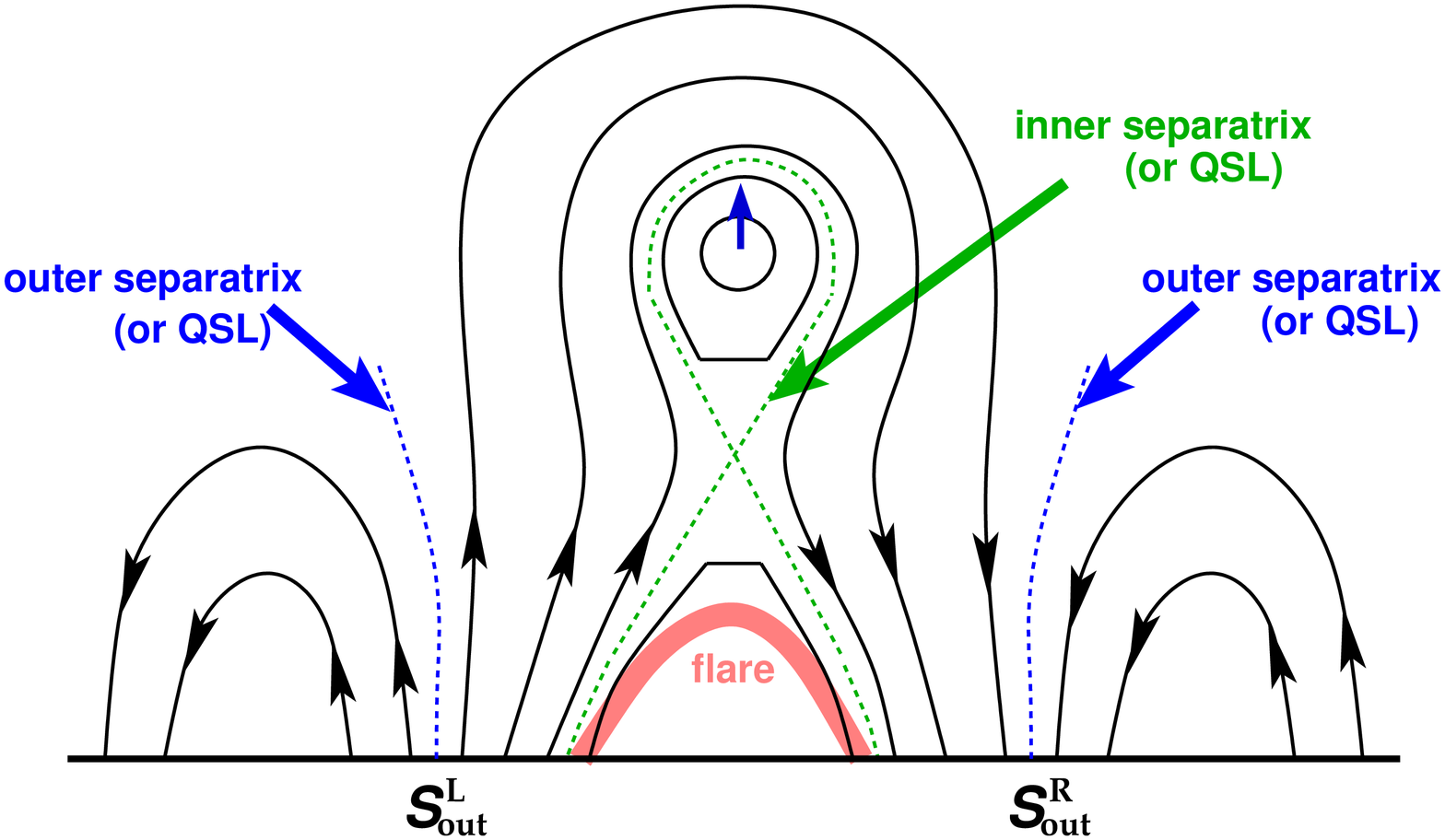}}
\noindent{\footnotesize {\bf Figure 1}\quad A sketch of magnetic reconnection
model with both inner and outer separatrices (or quasi-separatrix layers, QSLs)
being considered, where the black solid lines are magnetic field lines. It is
proposed in this paper that the flare ribbons would finally stop at the
intersection of the outer separatrices (or QSLs) with the
solar surface, i.e., $S_{out}^{L}$ and $S_{out}^{R}$.}

\begin{multicols}{2}

In the standard flare model, as a filament (or a magnetic flux rope in a general
sense) erupts, the overlying field lines are stretched up, leading to the
formation of a current sheet below the flux rope. As reconnection is triggered
and goes on in the current sheet, the reconnected field lines below the
reconnection area, along with the heated plasmas, pile up. The previously-heated
loops cool down due to radiation and heat conduction. Accelerated around
the reconnection area, energetic particles, along with thermal conduction, are
transferred down along the separatrix or quasi-separatrix layer (QSL) to heat
the chromosphere, forming H$\alpha$ ribbons and plasma evaporation. These
processes result in the typical observed features of two-ribbon flares, i.e.,
the apparent expansion of the flaring loop and the separation motion of the
flare ribbons (e.g., \cite{chen99a}). Such processes can keep going if the
magnetic field lines straddling over the flux rope extend to a long distance in
the horizontal direction, i.e., in the case of a large-scale bipolar field.
However, at least two factors may terminate such an on-going reconnection
process. One is that the reconnected field lines pile up to reach the
reconnection area, which then hinders the anti-parallel field lines from further
reconnecting. This factor can account for compact flares as demonstrated by
Chen et al. \cite{chen99}, but not for two-ribbon flares, where the current sheet extends up
well above the flare loop. The second factor, which we propose to account for
the limited lifetime of two-ribbon flares, is the existence of outer magnetic
separatrices or QSLs.

\quad  The idea is explained in Figure 1, the central part of which is
essentially the same as the CSHKP model, i.e., a flux rope with a null point
below resides inside a filament channel. An inner magnetic separatrix ({\it
green dashed line}) runs across the X-type null point. Note that in 3-dimensions the
null point is generalized to a quasi separator (or hyperbolic flux tube,
\cite{tito02}) and the inner separatrices are QSLs (which include separatrices
as a special case) \cite{prie92}. The difference of Figure 1 from the classical
CSHKP model is that there exist two outer magnetic separatrix (or QSL in
3-dimensions) segments on the two sides of the filament channel respectively, as
indicated by the blue dashed lines above $S_{out}^L$ and $S_{out}^R$. Outside
$S_{out}^L$--$S_{out}^R$ the field lines belong to different flux systems, which
can either be open field (i.e., a coronal hole) or closed field. As the flux
rope erupts, only the field lines straddling over the flux rope between
$S_{out}^L$ and $S_{out}^R$ can be stretched up and experience reconnection
below the flux rope. This means that the moving flare ribbons will finally stop
at the intersections of the outer magnetic separatrices (or QSLs) with the solar
surface, i.e., at $S_{out}^L$ and $S_{out}^R$.

\quad To confirm such a conjecture, we analyze the 2003 May 29 flare event, and study
the spatial relation between QSLs and the final positions of the flare ribbons.

\section{Observations and data analysis}

On 2003 May 29, a GOES X1.2-class flare occurred at S06W37 in the active region
AR10365. The flare started at $\sim$00:51 UT and peaked at 01:05 UT. UV ribbons
were almost invisible after $\sim$02:00 UT. It was a typical long-duration event,
showing two ribbons separating gradually. The flare loops and ribbons were well
observed by the Transition Region and Coronal Explorer ({\it TRACE},
\cite{hand99}) with a high spatial resolution of $1^{\prime\prime}$ and a cadence
of $\sim$3 min. The photospheric vector magnetograms across the flare were
obtained in Huairou Solar Observing Station ({\it HSOS}, \cite{ai87}) with a
pixel size of $0.35^{\prime\prime}$ and a cadence of $\sim$10 min. The
coalignment of the two datasets is accomplished with the help of the magnetogram
Michelson Doppler Imager (MDI) aboard the {\em Solar and Heliospheric
Observatory} ({\em SOHO}).

\end{multicols}

\centerline{\includegraphics[width=17cm]{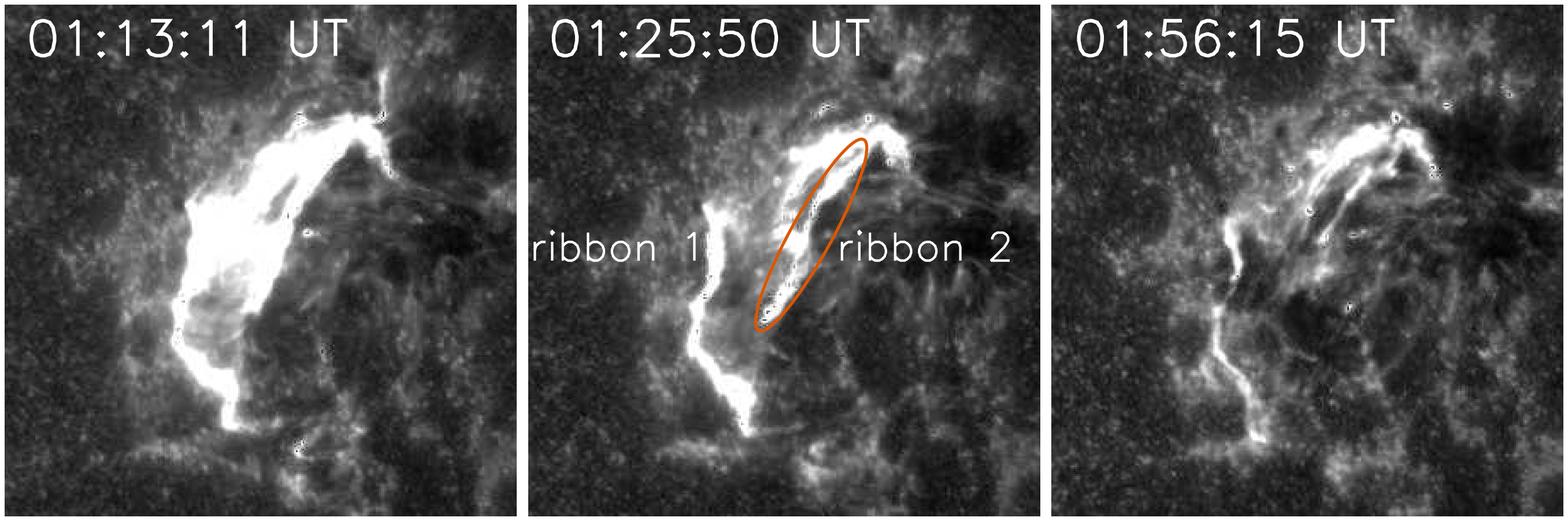}}
\noindent{\footnotesize {\bf Figure 2}\quad Evolution of the 2003 May 29 solar flare
in the {\em TRACE} 1600 {\AA} wavelength showing the separation of two ribbons. North
is up. The ribbons 1 and 2 are marked, and it is noted a bright patch to the north of
ribbon 2 persisted during the flare.}

\begin{multicols}{2}

\quad Figure 2 depicts the evolution of the flare in {\it TRACE} 1600 {\AA}, where two
ribbons were observed to separate slowly. Ribbon 1 is separate from any other
brightenings, whereas ribbon 2, marked by an ellipse, is connected to a bright
patch in the north. At 01:56:15 UT, the flare ribbons nearly approached their
final positions before fading away. Note that the bright patch to the north of
ribbon 2 persisted during the flare, whose nature is beyond the scope of this
paper.

\quad The coronal magnetic field is extrapolated from the {\it HSOS} vector
magnetogram before the flare peak, at 00:59 UT, with the non-linear force-free
model \cite{wieg04}. Figure 3 shows the extrapolated coronal
magnetic field lines with the photospheric magnetogram being rendered at
the bottom. The different magnetic flux systems are clearly identified, and the
boundaries between neighboring flux systems correspond to QSLs, across which
field lines go divergently. The flaring ribbons at 01:56:15 UT ({\it white
lines}) are located near the boundaries.

\vspace{5mm}
\centerline{\includegraphics[width=9cm]{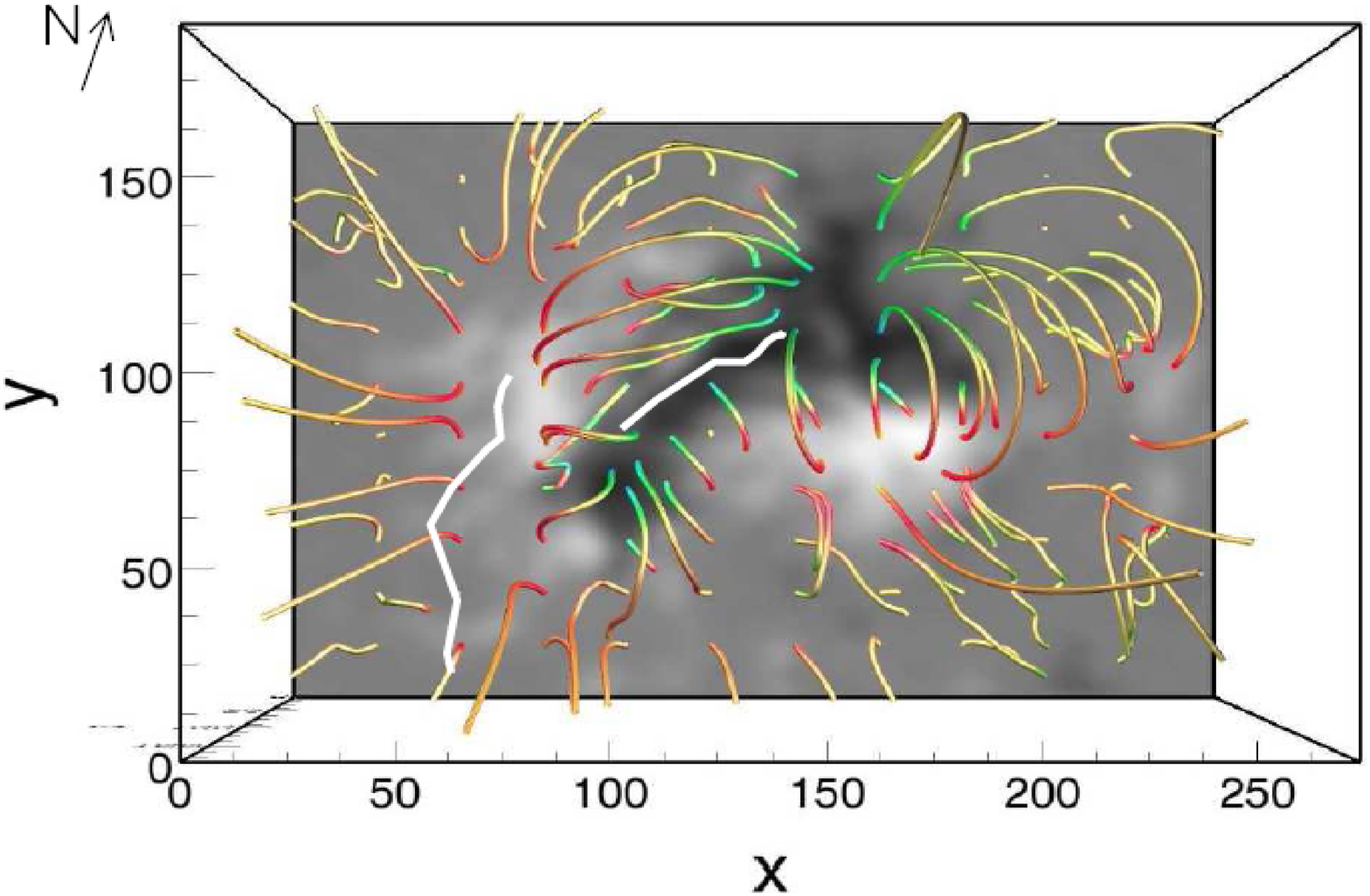}}
\noindent{\footnotesize {\bf Figure 3}\quad Top view of the extrapolated coronal
magnetic field ({\em solid lines}) that is anchored to the photospheric
magnetograms ({\em gray-scale}). The two white thick lines mark the locations
of the two flare ribbons. North is indicated by the arrow at the top-left
corner.}
\vspace{5mm}

\quad In order to compare the locations of the QSLs with those of the two
ribbons more quantitatively, we calculate the squashing degree $Q$, which
characterizes the magnetic connectivity, with $Q>>2$ corresponding to
a QSL \cite{demo06}. $Q$ is infinite at separatrices in theory, and is a very
large value due to finite size of the numerical grid. The $Q$-map ({\it yellow
lines}) is superimposed over the {\it TRACE} 1600 {\AA} intensity map at
01:56:15 UT in Figure 4. It can be seen that ribbon 2 near the end of the flare
is almost exactly cospatial with the intersection of the QSL at the solar 
surface. Although ribbon 1 is also roughly cospatial with the intersection of
the QSL, it is inclined with the QSL intersection with an angle of $20^\circ$.
Note that no null point or bald patch exists in the modeled box, we are not
sure whether the QSLs are separatrices.

\centerline{\includegraphics[width=8.5cm]{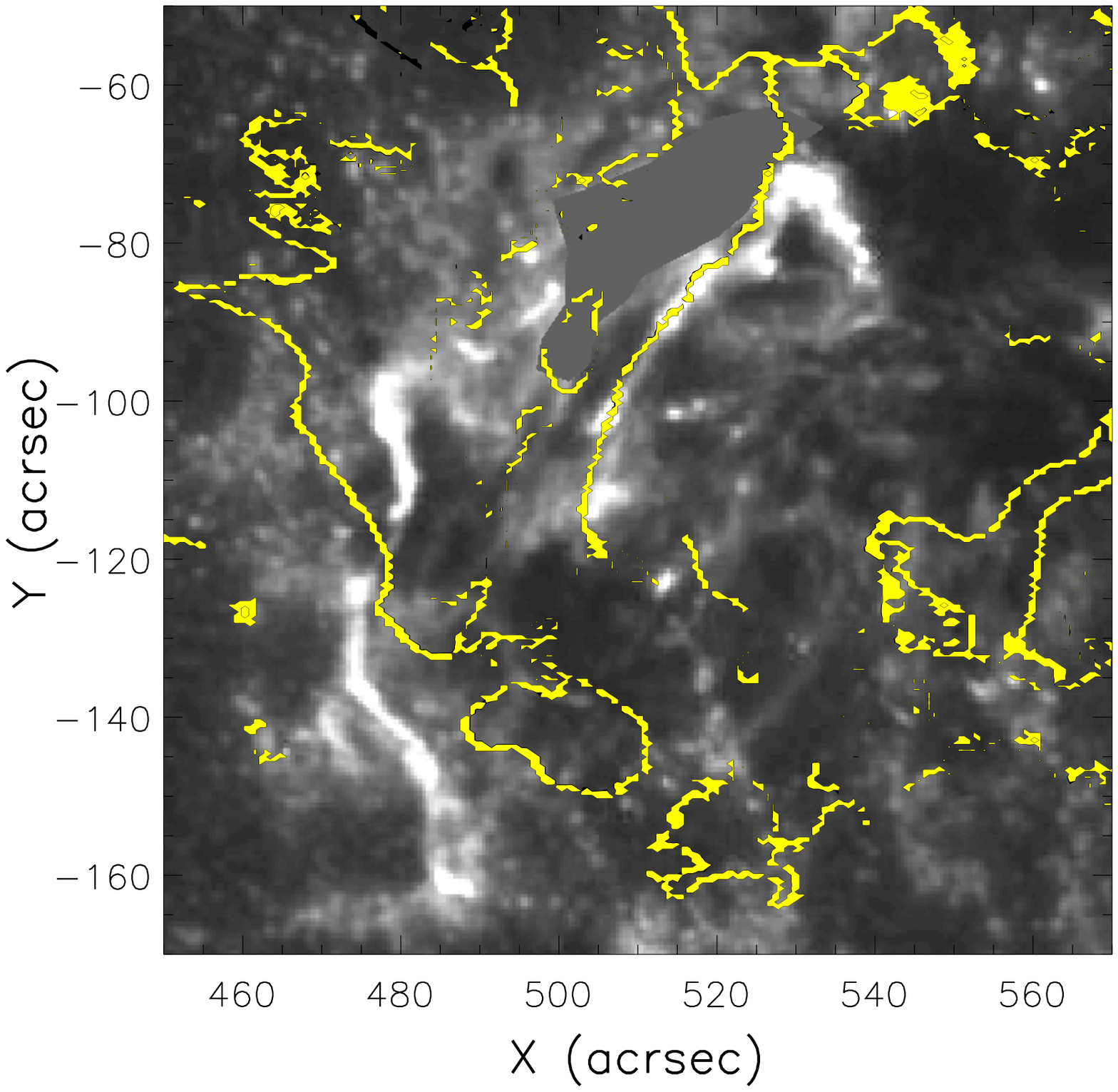}}
\noindent{\footnotesize {\bf Figure 4}\quad Comparison between the final
locations of the flare ribbons at 01:56:15 UT ({\em gray-scale}) with the
$Q$-map ({\em yellow lines} corresponding $Q=3100$) at 00:59 UT, which
indicates that the final location of ribbon 2 is cospatial with the footpoints
of a magnetic separatrix and that of ribbon 1 is roughly cospatial with the
footpoints of separatrix. North is up. Note that the bright patch to the north
of ribbon 2 is removed.}

\section{Discussions}

Magnetic QSLs play an important role in active region heating \cite{wang00} and
magnetic reconnection as well \cite{somo85, long01}. Solar flares, either
compact or two-ribbon flares, are widely explained in terms of magnetic
reconnection \cite{shib99}. In the reconnection model, magnetic
QSLs (or separatrices as a special case) divide the reconnected field
lines from the pre-reconnection field lines. Energetic particles and heat
conduction are transported down to evaporate the chromospheric plasma to form
flare loops. As the reconnection goes on, larger flaring loops are formed on
top of previous ones, with the two footpoints (two ribbons in 3-dimensions)
separating continually. According to such a reconnection model, at any time
during a flare, the flare ribbons, i.e., the footpoints of the flare loops, are
located at the intersection of this inner magnetic separatrix (or QSL) with the
solar surface, as illustrated by Figure 1. As the reconnection proceeds, this
inner separatrix (or QSL) moves outward horizontally along with the flare
ribbons. With this paper, we point out that there should also exist outer
separatrices (or QSLs) which border the filament channel, as marked by
$S_{out}^L$ and $S_{out}^R$ in Figure 1. After all the field lines between the
inner separatrices (or QSLs) and the outer separatrices (or QSLs) have
reconnected, no more field lines are available for further reconnection, and
magnetic reconnection is expected to halt. When this happens, the flare ribbons
reach the outer separatrices (or QSLs). In this paper, we analyzed the 2003 May
29 flare event, and found that the final location of ribbon 2 well matches
the outer QSL that is derived from the pre-flare magnetic field. Ribbon 1, however,
only roughly matches the outer QSL. The probable reason for the slight
discrepancy of ribbon 1 is that the field of view of the {\em HSOS} vector
magnetogram is too small and ribbon 1 is very close to the edge of the field of
view. With full-disk vector magnetograms, such a problem will be solved.

\quad Flare kernels and ribbons were often found to be almost cospatial with
the intersection of separatrices or QSLs with the solar surface \cite{gorb88,
mand91, demo94, vand94, schm97, mass09, su09}. Those authors related the ribbons
to the reconnection area, which tends to be a magnetic null point or QSL. In
their works, they generally picked up the flare ribbon images during the flare
process. According to the reconnection model, flare ribbons should be located at
the footpoints of the inner separatrix (or QSL). However, we stress that the
inner separatrix (or QSL), which is directly linked to the reconnection area,
might be difficult to derive with the current techniques of magnetic field
extrapolation. The derived QSLs in this paper, and in some of the previous
works, actually correspond to the outer separatrices (or QSLs), whose footpoints
are cospatial with the final location of flare ribbons. If we compare the flare
ribbon at any time with the derived QSLs, the two might always be roughly
cospatial, since in most flares the moving distance of a flare ribbon is only
$\sim 10^{\prime\prime}$ (e.g., \cite{wang03}), which is of the order of the
spatial resolution of many previous telescopes. With the unprecedented
resolutions of both vector magnetograms and UV images by {\it Solar Dynamics
Observatory} ({\em SDO}), more accurate comparisons between magnetic QSLs
and the final locations of flare ribbons will be very meaningful.

\quad The prediction of the flare occurrence is improving greatly. In this
paper we propose a theoretical conjecture to predict the final locations of
flare ribbons before the flare occurs. By assuming a suitable reconnection rate,
we can even further predict the lifetime of a flare before it occurs, which will
greatly enhance our capacity of space weather forecast.

\Acknowledgements{\bahao The authors thank P. D{\'e}moulin for helpful suggestions.
The research is supported by the Chinese foundations NSFC
(11025314, 10878002, and 10933003) and 2011CB811402. PFC is also supported by an
open research program of National Astronomical Observatories of China.}


\normalsize \parskip=0mm \baselineskip
18pt\renewcommand{\baselinestretch}{1.1}\footnotesize\parindent=4mm\bahao


\bibliographystyle{unsrt}

\end{multicols}

\
\
\vspace{2cm}
Please note that the page numbers have not been specified by the publishing
office. But the DOI number is correct.
\end{document}